\documentclass[aps,prl,twocolumn,showpacs,superscriptaddress]{revtex4}
\usepackage{graphicx}
\usepackage{amsmath,amssymb,bm}
\usepackage{times}
\usepackage[colorlinks,citecolor=blue,linkcolor=blue]{hyperref}

\begin{document}

\title{Geometric Heat Flux for Classical Thermal Transport in Interacting Open Systems}

\author{Jie Ren}\email{renjie@lanl.gov}
\affiliation{Department of
Physics and Centre for Computational Science and Engineering,
National University of Singapore, Singapore 117542, Republic of
Singapore}
\affiliation{NUS Graduate School for Integrative Sciences and
Engineering, Singapore 117456, Republic of Singapore}
\affiliation{Theoretical Division, Los Alamos National Laboratory, Los Alamos, New Mexico 87545, USA}

\author{Sha Liu}
\affiliation{NUS Graduate School for Integrative Sciences and
Engineering, Singapore 117456, Republic of Singapore}
\affiliation{Department of
Physics and Centre for Computational Science and Engineering,
National University of Singapore, Singapore 117542, Republic of
Singapore}

\author{Baowen Li}\email{phylibw@nus.edu.sg}
\affiliation{Department of
Physics and Centre for Computational Science and Engineering,
National University of Singapore, Singapore 117542, Republic of
Singapore}
\affiliation{NUS Graduate School for Integrative Sciences and
Engineering, Singapore 117456, Republic of Singapore}
\affiliation{NUS-Tongji Center for Phononics and Thermal Energy Science and Department of Physics, Tongji University, 200092 Shanghai, China}

\date{\today}

\begin{abstract}
We study classical heat conduction in a dissipative open system composed of interacting oscillators. By exactly solving a twisted Fokker-Planck equation which describes the full counting statistics of heat flux flowing through the system, we identify the geometric-phase-like effect and examine its impact on the classical heat transport. Particularly, we find that the nonlinear interaction as well as the closely related temperature-dependence of system-parameters are crucial in manifesting the geometric-phase contribution of heat flux. 
Finally, we propose an electronic experiment based on $RC$ circuits to verify our theoretical predictions.
\end{abstract}

\pacs{ 05.60.Cd, 03.65.Vf, 66.70.-f, 05.10.Gg}
\maketitle

Understanding the general features of transports is one of the
main goals in non-equilibrium statistical physics. Among many others, time-dependent driven transports like driven particle (mass, probability) transport and driven heat conduction are attracting an increasing attention. In particular, the latter
is of special interests \cite{TDHT3, TDHT1, TDHT2, TDHT22, TDHT23, TDHT32, TDHT4, TDHT5, Ren} because of both its theoretical and practical
importance in phononics \cite{Li_review}, where one may utilize temporal modulations
to alternatively achieve flexible dynamic control of thermal energy in
various phononic devices \cite{Li_review}.


In driven quantum systems, an intrinsic geometric contribution of the phase of a wave function will be emergent \cite{GePh} and it has been proved to have profound effects on many physical properties including thermal related ones \cite{GePh1, GePh2, GePh3}.
Similar geometric contributions have been also uncovered in the driven transport of noninteracting particles \cite{Parrondo1998, Shi2002, Astumian}.  These pioneering efforts culminated in the discovery of geometric-phase-like contribution in generating functions \cite{Sinitsyn, Sinitsyn_review}, which then inspired the identification of geometric heat flux in a single quantum junction \cite{Ren}. Namely, even under slow modulations, heat flux and fluctuations are not merely a simple temporal average of their static counterparts, but contain an extra geometric contribution regardless of driving rates.

Unlike noninteracting particle transport, heat conduction in solid is typically modeled by interacting lattices with reservoirs, where energy is transported in the absence of particle flow \cite{ADhar_review}. Therefore, whether and how the geometric contribution can emerge in classical heat conduction is still an open question. Moreover, the nonlinearity (anharmonicity) has been found of special importance in phononic devices \cite{Li_review}. However, the role of nonlinear interaction in the manifestation of geometric heat flux is still not yet explored, although works about time-dependent classical heat conduction in interacting lattices have already been carried out \cite{TDHT3, TDHT2, TDHT22, TDHT23, TDHT32, TDHT4}.

In this Letter, we shall address the above mentioned
objectives by exactly solving a twisted Fokker-Planck
equation, which describes the full counting statistics of heat flux flowing through a classical open system of interacting oscillators. We identify the geometric-phase effect on generating functions and examine its impact on the classical heat transport. In particular, we find that the nonlinearity of interaction as well as the related temperature-dependence of system parameters are crucial to the manifestation of geometric heat flux. Otherwise, for a linear system without temperature-dependent parameters, the geometric-phase effect is absent or only observable in high order heat fluctuations. Furthermore, by pointing out the analogy of a coupled $RC$ electric circuit and interacting oscillators, we are able to implement an electric experiment to verify the theoretical predictions of geometric-phase effects in heat transport.

\begin{figure}[t]
\scalebox{0.26}[0.24]{\includegraphics{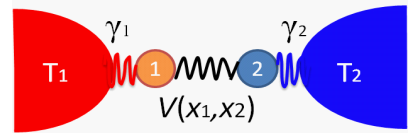}}
\vspace{-.5cm}
\caption{(color online). A sketch of two coupled classical Brownian oscillators in contact with two Langevin heat baths.}
\label{fig:scheme}
\end{figure}

We start with a typical interacting open system: two coupled Brownian oscillators in contact with two heat baths, as shown in Fig.~\ref{fig:scheme}. The vibrational dynamics is described by a set of Langevin equations:
$m_i\ddot{x}_i=-{\partial_{ x_i} V(x_1,x_2)}-\gamma_i \dot{x}_i+\xi_i$, $(i=1,2)$,
where $m_i$ and $x_i$ are the mass and displacement of oscillator $i$. $V(x_1, x_2)$ denotes the interaction potential. $\gamma_i$ depicts the viscosity of bath $i$, or say, the coupling strength between oscillator $i$ and bath $i$.
$\xi_{i}$ is the white noise with variance $\langle\xi_{i}(t)\xi_{j}(t')\rangle=2\gamma_{i}T_{i}\delta_{ij}\delta(t-t')$, where $T_i$ is the temperature of bath $i$. $Q(t)=\int^t_0{\partial_{x_1}V(x_1,x_2)}\dot{x}_1dt'$ is defined to describe the heat transferred from heat bath $1$ to $2$ during time $t$. This vibrational dynamics is a typical interacting model for heat transport which has been also used to describe the well-known Feynman ratchet-pawl model \cite{ratchet}.
For microdynamics at low Reynolds number where oscillators either possess
nano-sizes or move in an extremely viscous media, the inertia terms are
negligible \cite{Purcell}, and in turn we have an overdamped dynamics:
\begin{equation}
\begin{array}{l}
\gamma_1 \dot{x}_1+{\partial_{ x_1} V(x_1,x_2)}=\xi_1, 
\\
\gamma_2 \dot{x}_2+{\partial_{ x_2} V(x_1,x_2)}=\xi_2.  
\end{array}\label{overdamped}
\end{equation}
For an harmonic coupling $V(x_1,x_2)=k(x_1-x_2)^2/2$ with spring constant $k$, the system can be described by a Fokker-Planck equation \cite{FP}:
\begin{equation}\label{eq:FP}
\frac{\partial\rho(y,t)}{\partial
t}=\bigg[\left(\frac{T_1}{\gamma_1}+\frac{T_2}{\gamma_2}\right)\frac{\partial^2}{\partial{y^2}}
+\left(\frac{k}{\gamma_1}+\frac{k}{\gamma_2}\right)\frac{\partial}{\partial
y} y\bigg]\rho(y,t),
\end{equation}
where $y(t)=x_1(t)-x_2(t)$. Then the transferred heat from bath $1$ to $2$ within time $t$ is $Q(t)=\int^t_0dt'ky(\xi_1-ky)/{\gamma_1}$.

To study the time-dependent driven heat transport, we introduce the characteristic function of the joint probability $\rho(y,Q,t)$, defined as
$z(y,\chi,t)=\int^{\infty}_{-\infty}dQe^{\bm{i}\chi Q}\rho(y,Q,t)$.
This characteristic function satisfies a twisted Fokker-Planck equation:
\begin{eqnarray}\label{eq:twistFP}
\partial_t z(y,\chi,t)=L_{\chi}(t)z(y,\chi,t),
\end{eqnarray}
where, following derivations from~\cite{Visco1, LD_harmonic}, one can find,
\begin{equation}
\begin{array}{l}
L_{\chi}=\left(\frac{T_1}{\gamma_1}+\frac{T_2}{\gamma_2}\right)\frac{\partial^2}{\partial
y^2}
+k\left(\frac{1}{\gamma_1}+\frac{1}{\gamma_2}-\frac{2T_1}{\gamma_1}\bm{i}\chi\right)y\frac{\partial}{\partial
y} \\
+k^2\left(\frac{T_1}{\gamma_1}\left(\bm{i}\chi\right)^2-\frac{1}{\gamma_1}\bm{i}\chi\right)y^2+k\left(\frac{1}{\gamma_1}+\frac{1}{\gamma_2}-\frac{T_1}{\gamma_1}\bm{i}\chi\right).
\label{eq:Lx}
\end{array}
\end{equation}
The twisted Fokker-Planck operator $L_{\chi}(t)$ is generally time-dependent,
wherein parameters $k(t)$, $\gamma_j(t)$ and $T_j(t)$ could be subject to adiabatically cyclic modulations. The ``adiabatic'' here means the modulation period $T_p$ is much larger than the system's characteristic time scale of relaxation, $T_c$, namely \cite{supply}:
\begin{equation}
T_p\gg T_c=\frac{\gamma_1\gamma_2}{k(\gamma_1+\gamma_2)}.
\label{eq:adiabatic}
\end{equation}

It is easy to verify that Eqs. (\ref{eq:twistFP}) and (\ref{eq:FP}) have the same initial condition $z(y,\chi,0)=\rho(y,0)$. When $\chi=0$, we have $z(y,0,t)=\rho(y,t)$ from the definition and in turn Eq. (\ref{eq:twistFP}) reduces to Eq. (\ref{eq:FP}).
Integrating over the degree of freedom $y$ in $z(y,\chi,t)$, we obtain the characteristic
function of $Q$:
$Z(\chi,t)=\int dy z(y,\chi,t)
=\int dQ e^{\bm{i}\chi Q}P(Q,t),$
with $P(Q,t)=\int dy\rho(y,Q,t)$.
Thus, the cumulant generating function is
$G(\chi)=\lim_{t\rightarrow\infty}t^{-1}\ln{Z(\chi,t)}$, which generates the $n$-order cumulant of heat fluctuations: $\lim_{t\rightarrow\infty}\langle\langle Q^n\rangle\rangle/t=\partial^n_{\bm{i}\chi} G(\chi)|_{\chi=0}$.

Following \cite{Sinitsyn}, the
cumulant generating function of adiabatically driven system can be separated into two parts --the dynamic contribution and the geometric one:
$G(\chi)=\lim_{t\rightarrow\infty}\frac{1}{t}\ln Z(\chi,t)=G_{\mathrm{dyn}}+G_{\mathrm{geom}}$.
The dynamic contribution, $G_{\mathrm{dyn}}=\frac{1}{T_p}\int^{T_p}_0dt\lambda_0(\chi,t)$, with $\lambda_0(\chi,t)$ denoting the ground-state eigenvalue of $L_{\chi}(t)$, survives whenever system parameters are static, or experience single or multiple modulations; whereas the appearance of the geometric contribution $G_{\mathrm{geom}}$ requires at least two parameter modulations. For the case of periodically driving pairs $(u_1(t), u_2(t))$, which could be chosen from any two of $k, \gamma_j, T_j$, we have \cite{supply}
\begin{align}
G_{\mathrm{geom}}&=-\frac{1}{T_p}\iint_{u_1u_2}du_1du_2
\mathcal{F}_{u_1u_2}(\chi), \\
\text{with }&\mathcal{F}_{u_1u_2}(\chi)\equiv\int^{\infty}_{-\infty} dy \left[\frac{\partial\varphi_0}{\partial u_1}\frac{\partial\psi_0}{\partial u_2}-\frac{\partial\varphi_0}{\partial u_2}\frac{\partial\psi_0}{\partial u_1}\right],
\label{Fcur}
\end{align}
where $\varphi_0(\psi_0)$ denotes the corresponding left(right) ground-state eigenfunction, and the subscript $u_1u_2$ denotes the integral area enclosed by the modulating contour. Clearly, $\mathcal{F}_{u_1u_2}(\chi)$ has the physical meaning of curvature of the parameter space $(u_1, u_2)$ for $L_{\chi}$. It is of pure geometric origin and is independent of the modulation speed. Mathematically, $G_{\mathrm{geom}}$ is an analog of the geometric Berry phase \cite{Sinitsyn_review} in quantum mechanics, where the wave function will gain an extra phase after a cyclic evolution \cite{Berry,Garrison88}. Similarly, in the full counting statistics of cyclic driven systems, the cumulant generating function (analog of phase) in the exponent of the characteristic function (analog of wave function) will also gain an additional term
\cite{Sinitsyn,Sinitsyn_review}. Both extra terms share the similar geometric origin from the nontrivial curvature in the system's parameter space.
In turn the $n$th cumulant of heat fluctuations has two separate contributions as well~\cite{Sinitsyn, Ren}:
\begin{eqnarray}
\lim_{t\rightarrow\infty}\frac{\langle\langle Q^n\rangle\rangle}{t}=\left.\frac{\partial^nG_{\mathrm{dyn}}}{\partial(\bm{i}\chi)^n}\right|_{\chi=0}+
\left.\frac{\partial^nG_{\mathrm{geom}}}{\partial(\bm{i}\chi)^n}\right|_{\chi=0}.
 \end{eqnarray}
Of prime interest is the first cumulant, the average heat flux $J=J_{\mathrm{dyn}}+J_{\mathrm{geom}}$, with
\begin{eqnarray}
{J}_{\mathrm{dyn}}&=&\frac{1}{T_p}\int^{T_p}_0dt \left.\frac{\partial \lambda_0(\chi,t)}{\partial(\bm{i}\chi)}\right|_{\chi=0},   \label{Jd}\\
{J}_{\mathrm{geom}}&=&-\frac{1}{T_p}\iint_{u_1u_2}du_1du_2
\left.\frac{\partial {\mathcal{F}}_{u_1u_2}(\chi)}{\partial(\bm{i}\chi)}\right|_{\chi=0}. \label{Jg}
\end{eqnarray}

Apparently, to study the geometric heat flux as well as other interest transport properties, we need to solve the eigen-problem of $L_{\chi}$, which however spans the infinite-dimensional Hilbert space and is usually difficult to tackle. Fortunately, after some algebra, we can cast Eqs.~(\ref{eq:twistFP}, \ref{eq:Lx}) to the Schr\"odinger's eigen-problem of quantum harmonic oscillator
\cite{supply}, with the help of which we analytically obtain 
the ground-state eigenvalue for $L_{\chi}(t)$,
\begin{equation}
\lambda_0(\chi,t)={r}\left(1-\theta\right)/2,
\label{lambda}
\end{equation}
and the right ground-state eigenfunction,
\begin{equation}
\psi_0(y,\chi,t)=\exp\left({-\frac{r+r\theta-2\bm{i}\chi kD_1}{4D}y^2}\right), \label{right}
\end{equation}
as well as the corresponding left eigenfunction,
\begin{equation}
\varphi_0(y,\chi,t)=\sqrt{\frac{r\theta}{2\pi
D}}\exp\left({\frac{r-r\theta-2\bm{i}\chi kD_1}{4D}y^2}\right),
\label{left}
\end{equation}
where $\theta=\sqrt{1-{4k^2D_1D_2}/{r^2}\left(-\chi+\bm{i}({1}/{T_2}-{1}/{T_1})
\right)\chi}$, $r=r_1+r_2$, $D=D_1+D_2$ with $r_{j}=k/\gamma_{j}$
and $D_{j}=T_{j}/\gamma_{j}$.

Substituting Eq.~(\ref{lambda}) into Eq.~(\ref{Jd}), we see that the dynamic heat flux is just the temporal average of its static counterpart: $J_{\mathrm{dyn}}=T_p^{-1}\int^{T_p}_0dt J_{\mathrm{st}}(t)$, with $J_{\mathrm{st}}\equiv\partial_{\bm{i}\chi}
\lambda_{0}|_{\chi=0}={k(T_1-T_2)}/{(\gamma_1+\gamma_2)}$, where the integral concerns all possible time-dependent parameters. It is interesting to notice that generally given $T_1(t)=T_2(t)$, not only the average, but all the odd order cumulants of the dynamic flux will vanish as well, due to the even symmetry $\lambda_0(\chi)\equiv\lambda_0(-\chi+\bm{i}(1/T_2-1/T_1))=\lambda_0(-\chi)$.

Substituting Eqs.~(\ref{right}, \ref{left}) into Eq.~(\ref{Fcur}) and in turn Eq.~(\ref{Jg}), we can then study the geometric heat flux ${J}_{\mathrm{geom}}$ under any pair-parameter manipulations. We first consider a special
case of two system-bath couplings $\gamma_1(t), \gamma_2(t)$ being modulated. In this case, we find the curvature $\mathcal{F}_{\gamma_1\gamma_2}\equiv0$, that is, 
no matter how arbitrarily one drives these two couplings, the geometric contributions to all cumulants of heat transport are
always zero. The geometric effect is absent in the case of merely modulating system-bath couplings. Similar phenomenon is also observed in quantum heat
transport \cite{Ren}, which implies that this may have connection with some universal pumping restrictions \cite{GUC} in open systems. Furthermore, this absence of geometric heat flux may relate to a no-pumping theorem of a different quantity, probability current, in closed driven systems without explicitly connecting with heat baths \cite{Horowitz}. It shows that, in terms of our Eq.~(\ref{eq:FP}), the probability current is absent when $\frac{k(1/\gamma_1+1/\gamma_2)}{(T_1/\gamma_1+T_2/\gamma_2)}$ is time-independent. In fact, when $T_1=T_2$, this ratio is indeed independent of $\gamma_i$ so that there is no-pumping of probability current no matter how we drive $(\gamma_1, \gamma_2)$.

In the case of modulating any other combination of two system parameters, 
however, the geometric
contribution emerges. For example, if we modulate one system-bath coupling $\gamma_2(t)$
and one bath temperature $T_1(t)$, 
the nonzero first derivative of curvature, $\partial_{\bm{i}\chi}\mathcal{F}_{\gamma_2T_1}|_{\chi=0}=-{\gamma_1\gamma_2}/[2(\gamma_1+\gamma_2)^3]\neq0$ can induce nonzero geometric heat flux $J_{\mathrm{geom}}\neq0$. For another example, if two baths are kept isothermal $T_1=T_2=T_0$, then the dynamic heat flux will be always zero, $J_{\mathrm{dyn}}=0$. Even so, we still can realize the nonzero heat transport through the geometric contribution by modulating, like $(k,
\gamma_2)$, so that
$J=J_{\mathrm{geom}}={T_p^{-1}}\iint_{k,\gamma_2}dk
d\gamma_2 \left\{{\gamma_1T_0}/[{2k(\gamma_1+\gamma_2)^2}]\right\}$.

The most typical modulation is driving bath temperatures $(T_2, T_1)$ \cite{TDHT2,TDHT22,TDHT23,TDHT32}, because implementing this protocol is much easier than modulating $k(t)$ and $\gamma_j(t)$ in practice. The curvature for this two-temperature driving reads:
\begin{equation}
\mathcal{F}_{T_2T_1}=\frac{\gamma_1\gamma_2(\gamma_2-\gamma_1)(\bm{i}\chi)^2}{2(\gamma_1+\gamma_2)^3\theta^3}.
\end{equation}
When the couplings are symmetric, $\gamma_2=\gamma_1$, one can see
that $\mathcal{F}_{T_2T_1}\equiv0$. That is, the geometric-phase effect is absent in the symmetric linear (harmonic) system. Interestingly, even when two couplings are asymmetric, $\gamma_2\neq\gamma_1$, we find
$\partial_{\bm{i}\chi}\mathcal{F}_{T_2T_1}|_{\chi=0}=0$
in spite of nonzero $\mathcal{F}_{T_2T_1}$, so that the geometric heat flux is still vanishing, $J_{\mathrm{geom}}\equiv\partial_{\bm{i}\chi}
G_{\mathrm{geom}}|_{\chi=0}=0$. In other words, although
the geometric effect exists in the classical asymmetric linear system, its contribution is not observable if one measures only the average flux. The geometric effect can only manifest itself in the higher order heat fluctuations, like the shot noise of currents, i.e., $\partial^2_{\bm{i}\chi}G_{\mathrm{geom}}|_{\chi=0}\propto\partial^2_{\bm{i}\chi}\mathcal{F}_{T_2T_1}|_{\chi=0}\neq0$.

Recalling the modulation $(\gamma_2, T_1)$
can induce the nonzero geometric heat flux, we speculate that as long as the viscosity $\gamma_2$ is temperature dependent, the modulation $(T_2, T_1)$  can be effectively equal to the modulation $(\gamma_2, T_1)$. As a consequence, in such systems of temperature-dependent viscosity, merely modulating $(T_2, T_1)$ may produce nonzero curvature thus nonzero geometric heat flux. In fact, viscosity is generally temperature-dependent \cite{Temp_viscosity}. Phenomenologically, we can assume
$\gamma_j=\gamma_0+a{T_j}^n$, with $j=1,2$.
Thus for the typical modulation $(T_2, T_1)$, we indeed obtain the nonzero geometric heat flux $J_{\mathrm{geom}}=$
\begin{align}
\frac{1}{T_p}\!\!\iint_{T_2 T_1}\!\!\!\!\!\!\!\!dT_2dT_1 \!
\frac{an
(\gamma_0+a{T_1}^n)(\gamma_0+a{T_2}^n)({T_1}^{\!n-1}\!\!+{T_2}^{\!n-1})}{2\left(2\gamma_0+a{T_1}^n+a{T_2}^n\right)^3}.
\label{nonlinear-berry}
\end{align}
Clearly, the temperature-dependence plays a key role for the manifestation of geometric effect. Given a modulation cycle, the geometric heat flux $J_{\mathrm{geom}}$ vanishes when the temperature-dependency $a\rightarrow0$; while $J_{\mathrm{geom}}$ increases to a saturated value as $a$ increases. When $n=0$, $\gamma_j$ also becomes temperature independent, and in turn $J_{\mathrm{geom}}$ becomes zero as well.

Considering the intrinsic effective temperature-dependencies of system parameters are ubiquitous in nonlinear interacting oscillators \cite{Li2007, SCPT}, we thus speculate that with the help of nonlinearity, the existing geometric effect is able to manifest itself into the geometric heat flux as well. Therefore, we further consider the 
FPU-$\beta$ model \cite{Li_review}, with nonlinear interacting potential:
$V(x_1, x_2)=\frac{k_1}{2}(x_1-x_2)^2+\frac{k_2}{4}(x_1-x_2)^4$.
As we shall see in follows, that the nonlinear strength $k_2$ can induce effective temperature dependence of system parameters~\cite{Li2007, SCPT} is the
crucial ingredient to manifest geometric heat flux.

\begin{figure}
\scalebox{0.35}[0.34]{\includegraphics{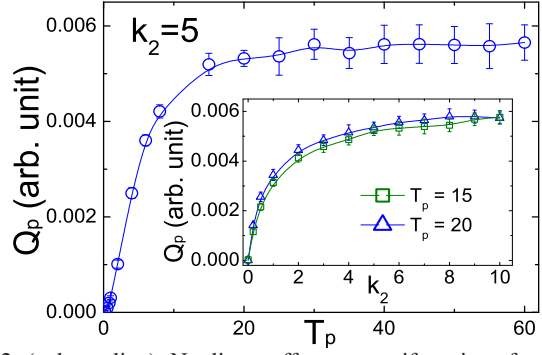}}
\vspace{-.5cm}
\caption{(color online). Nonlinear effect on manifestation of geometric effect in classical heat transport for FPU-$\beta$ model. $k_1 = 0.5$, $\gamma_1 = \gamma_2 = 5$. The physical value correspondences of those dimensionless units can be found in Ref.~\cite{Li_review}. The error bar denotes the standard derivation of $25$ times simulations, each is averaged over $10^6$ periods.}
\label{fig:nonlinear}
\end{figure}

In Fig.~\ref{fig:nonlinear}, we numerically simulate Eq.~(\ref{overdamped}) and calculate the average geometric heat $Q_{p}\equiv J_{\mathrm{geom}}T_p$, defined for a driving cycle. Given the temperature driving protocol: $T_2=0.09+0.06\cos({2\pi}t/{T_p}+\pi/4), T_1=0.09+0.06\sin({2\pi} t/{T_p}+\pi/4)$, we have a zero dynamic flux zero but a nonzero $Q_{p}$. When $T_p$ becomes large (adiabatic limit), $Q_p$ saturates to a fixed value independent of $T_p$, which indicates that it is purely a geometric property. The geometric heat per driving cycle does not rely on the driving rate, but only depends on the geometry of the driving contour in parameter spaces. The deviations of $Q_p$ from the fixed value at the fast driving regime are due to the breakdown of the adiabatic precondition.

The inset of Fig.~\ref{fig:nonlinear} verifies that when the nonlinear interaction reduces to harmonic coupling $(k_2=0)$, the geometric contribution disappears $(J_{\mathrm{geom}}=0)$. Only with the help of nonlinearity,
the geometric effect of temperature modulations can manifest itself into the heat flux. Moreover, increasing $k_2$ can enhance $Q_p$, until to a saturated value, which coincides qualitatively with the behavior of the analytic result Eq.~(\ref{nonlinear-berry}), by increasing $a$. 
In fact, from the viewpoint of nonequilibrium Green's functions \cite{wangEPJB}, the nonlinear interaction effect in thermal transport is reflected in the temperature-dependent effective self-energies, which in our case are exactly the temperature-dependent $\gamma_i$.

Although we focus on a two-coupled-oscillator system at the moment,
it could be straightforward to generalize the above analysis into arbitrarily long coupled-oscillator model with inertial terms, of which the eigenvalues and eigenvectors
of the twisted Fokker-Planck operator can be obtained in terms of appropriate phonon
Green's functions~\cite{LD_harmonic}.

Now, we shall examine the previous studies to see whether it is justified for neglecting geometric heat flux. In~\cite{TDHT2, TDHT22}, only one bath temperature is under cyclic manipulation so that the absence of geometric heat flux is justified, because for cyclic driving, the manifestation of geometric effect requires at least two parameter modulations in order to enclose a nonzero area in the parameter space. The same is true in~\cite{TDHT4}, where the interacting lattice is only cyclically driven by one mechanical force. In~\cite{TDHT22, TDHT23, TDHT32}, although two bath temperatures are modulated, they are varying either isothermally $T_1=T_2=T_0+\Delta T(t)$ \cite{TDHT23} or inversely $T_1=T_0+\Delta T(t)$, $T_2=T_0-\Delta T(t)$ \cite{TDHT22,TDHT32}. In this way, the modulation contour only closes as a line segment. Therefore, the absence of geometric heat flux is also justified.

\begin{figure}
\vspace{-.1cm}
\scalebox{0.33}[0.35]{\includegraphics{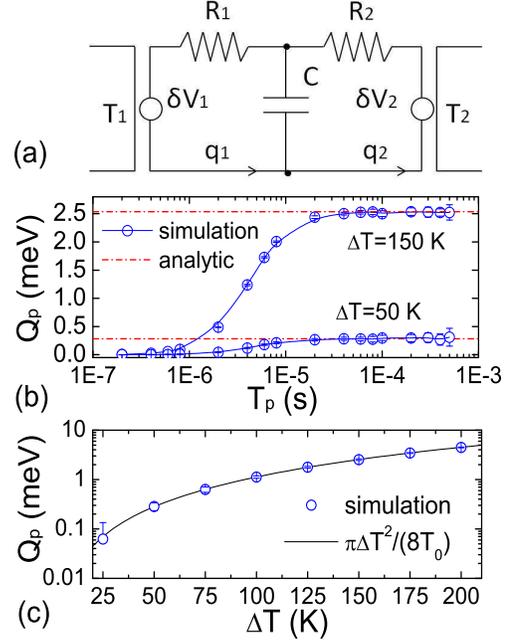}}
\vspace{-.4cm}
\caption{(color online). (a) A parallel $RC$ electric circuit subject to Johnson-Nyquist noises. The analogy with two coupled Brownian oscillators with heat baths is reflected by the parameter correspondences: ($q_j$, $R_j$, $1/C$, $\delta V_j$) $\leftrightarrow$ ($x_j$, $\gamma_j$, $k$, $\xi_j$). (b) Manifestation of geometric effect for classical heat transport in $RC$ electric circuit with temperature-dependent resistances. (c) $Q_p=J_{\mathrm{geom}}T_p$ as a function of $\Delta T$. The simulations are well fitted by ${\pi\Delta T^2}/{(8T_0)}$, approximated from Eq. (\ref{RC-berry}). Parameters are $a=10$k$\Omega$/K, $C=1$pF, $T_0=300$K.}
\label{fig:RC}
\end{figure}

In the last part, we would like to propose an experimental implementation to demonstrate our predictions on manifestation of geometric effect in classical heat transport through coupled oscillators. In view of the well-known electric analogy of interacting oscillators' Brownian motion \cite{RC,FP}, we are able to map the oscillator system into a $RC$ circuit, where two resistors of resistance $R_j$ are arranged in parallel with a capacitor of capacitance $C$, as shown in Fig.~\ref{fig:RC}(a). The left (right) circuit part is subjected to a thermal reservoirs of temperature $T_1$ ($T_2$), which generates a Gaussian voltage fluctuation $\delta V_1$ ($\delta V_2$), so called Johnson-Nyquist noise, with variance $\langle\delta V_i(t)\delta V_j(t')\rangle=2R_iT_i\delta_{ij}\delta(t-t')$  \cite{FP, RC}. 
$q_j$ denotes the charge going through the resistor $R_j$ and $dq_j/dt$ is the corresponding electric current. Consequently, the dynamics of this $RC$ circuit is described by:
\begin{equation}
\begin{split}
R_1 {dq_1}/{dt}+(q_1-q_2)/C=\delta V_1, 
\\
R_2 {dq_2}/{dt}+(q_2-q_1)/C=\delta V_2. 
\end{split}\label{RC}
\end{equation}
Note that Eq.~(\ref{RC}) is of a similar form as the overdamped dynamics Eq.~(\ref{overdamped}) for interacting oscillators.
The heat transferred over time $t$ now is analogously defined as $Q(t)=\int^t_0\dot q_1(q_1-q_2)/Cdt'$, which actually is the charging work done by the left reservoir on the capacitor. Thereafter, we can modulate $C(t), R_j(t), T_j(t)$ to test our predictions.

Nonlinear capacitor can be used to simulate the nonlinearity effect. To mimic the related temperature-dependent viscosity, we can choose resistors of temperature-dependent resistance, e.g. $R_j=aT_j$, with contacting them to their respective reservoirs. From Eq.~(\ref{nonlinear-berry}), we then have
\begin{align}
J_{\mathrm{geom}}
&=
\frac{1}{T_p}\iint_{T_2 T_1}dT_2dT_1
\frac{T_1T_2}{\left({T_1}+{T_2}\right)^3}.
\label{RC-berry}
\end{align}
For cyclic modulation: $T_2=T_0+\Delta T\cos({2\pi}t/{T_p}+\pi/4)$, $T_1=T_0+\Delta T \sin({2\pi} t/{T_p}+\pi/4)$, we will have zero $J_{\mathrm{dyn}}$ but nonzero $J_{\mathrm{geom}}$. Although this geometric heat flux is independent of $C$ and $R_j$, the adiabatic precondition Eq.~(\ref{eq:adiabatic}) requires $T_p\gg T_c=CR_1R_2/(R_1+R_2)$. Assume we use $a=10$ k$\Omega$/K, $C=1$ pF, $T_0=300$ K, we will have 
the adiabatic condition $T_p\gg T_c\approx3$ $\mu$s. Analytic results are verified by numerical simulations of Eq.~(\ref{RC}), as plotted in Fig.~\ref{fig:RC}(b). Clearly, for $T_p=20$ $\mu$s, the geometric heat per cycle already reaches the adiabatic limit. Therefore, as long as the system evolves a long time, i.e. with large number of cycles,  we can accumulate a large geometric heat, e.g., for $T_p=20$ $\mu$s, $\Delta T=50$ K, $Q_p=0.283$ meV, after each minute we can have $847$ eV. Moreover, if we scale up many $RC$ circuits in parallel, we are able to obtain even larger geometric heat flux. Alternatively, by increasing the modulation amplitude, one can also increase geometric heat flux, as shown in Fig. \ref{fig:RC}(c).

 Given the fact that the fluctuation theorems in electric circuits \cite{FTexp} as well as the energy rectification  have been experimentally realized in a nonlinear electrical transmission line \cite{Diodeexp}, we believe that our prediction of geometric-phase effect on heat transport can be also experimentally validated in a foreseeable future. Furthermore, as our ability to design and manipulate
nano/micro-sized systems improves, we believe that the present study of geometric energy (heat) flux could provide a new means of energy harvesting by harnessing the ubiquitous cyclic changes in the universe.

\begin{widetext}
\section{Supplementary Material for ``Geometric Heat Flux of Classical Thermal Transport in Interacting Open Systems''}

%
%
%
%
%
%

In this supplement, we are going to (1) analytically solve the eigen-problem of the twisted Fokker-Planck equation, (2) expose the condition of so-called ``adiabatic'', (3) detail the derivations of geometric-phase effect in generating functions, which finally leads to the geometric heat flux.

\subsection{Exact solution of the twisted Fokker-Planck equation}
Transport behaviors in the long time limit are of our central interest. They are
governed by the ground state of the twisted Fokker-Planck operator $L_{\chi}$. We thus need first exactly solve the eigen-problem of the twisted Fokker-Planck equation, with time-independent $L_{\chi}$:
\begin{eqnarray}
\partial_t z=L_{\chi}z,
\end{eqnarray}
with
\begin{equation}
L_{\chi}=k_B\left(\frac{T_1}{\gamma_1}+\frac{T_2}{\gamma_2}\right)\frac{\partial^2}{\partial
y^2}
+k\left(\frac{1}{\gamma_1}+\frac{1}{\gamma_2}-\frac{2k_BT_1}{\gamma_1}\bm{i}\chi\right)y\frac{\partial}{\partial
y}+k^2\left(\frac{k_BT_1}{\gamma_1}\left(\bm{i}\chi\right)^2-\frac{1}{\gamma_1}\bm{i}\chi\right)y^2+k\left(\frac{1}{\gamma_1}+\frac{1}{\gamma_2}-\frac{k_BT_1}{\gamma_1}\bm{i}\chi\right).
\end{equation}
Note that in the main text, we set $k_B=1$ for the sake of clarity. We make an ansatz of the solution:
\begin{equation}
z(y,\chi,t)=\sum_{n=0}^{\infty}C_ne^{\lambda_nt}e^{-\frac{y^2}{4T}}f_n(y),
\end{equation}
where $C_n$ is the coefficient depending on initial conditions, $\lambda_n$ denotes the $n$-th eigenvalue of $L_{\chi}$, $e^{-\frac{y^2}{4T}}f_n(y)$ is the corresponding eigenfunction, and $T$ is a parameter to be determined in follows. Put this ansatz into the twisted Fokker-Planck equation, and utilize the following relations:
\begin{eqnarray}
\frac{\partial}{\partial
y}e^{\frac{-y^2}{4T}}f_n(y)&=&-\frac{y}{2T}e^{\frac{-y^2}{4T}}f_n(y)+e^{\frac{-y^2}{4T}}\frac{\partial
f_n(y)}{\partial
y},\\
\frac{\partial^2}{\partial
y^2}e^{\frac{-y^2}{4T}}f_n(y)&=&-\frac{1}{2T}e^{\frac{-y^2}{4T}}f_n(y)+\frac{y^2}{4T^2}e^{\frac{-y^2}{4T}}f_n(y)-\frac{y}{2T}e^{\frac{-y^2}{4T}}\frac{\partial
f_n(y)}{\partial y}-\frac{y}{2T}e^{\frac{-y^2}{4T}}\frac{\partial
f_n(y)}{\partial
y}+e^{\frac{-y^2}{4T}}\frac{\partial^2f_n(y)}{\partial y^2},
\end{eqnarray}
we then have
\begin{multline}
\bigg\{k_B\left(\frac{T_1}{\gamma_1}+\frac{T_2}{\gamma_2}\right)\frac{\partial^2}{\partial
y^2}+\left[k\left(\frac{1}{\gamma_1}+\frac{1}{\gamma_2}-\frac{2k_BT_1}{\gamma_1}\bm{i}\chi\right)-k_B\left(\frac{T_1}{\gamma_1}+\frac{T_2}{\gamma_2}\right)\frac{1}{T}\right]y\frac{\partial}{\partial
y}\\
+\left[\frac{k_B}{4T^2}\left(\frac{T_1}{\gamma_1}+\frac{T_2}{\gamma_2}\right)-\frac{k}{2T}\left(\frac{1}{\gamma_1}+\frac{1}{\gamma_2}-\frac{2k_BT_1}{\gamma_1}\bm{i}\chi\right)+k^2\left(\frac{k_BT_1}{\gamma_1}\left(\bm{i}\chi\right)^2-\frac{1}{\gamma_1}\bm{i}\chi\right)\right]y^2+\\
k\left(\frac{1}{\gamma_1}+\frac{1}{\gamma_2}-\frac{k_BT_1}{\gamma_1}\bm{i}\chi\right)-\frac{k_B}{2T}\left(\frac{T_1}{\gamma_1}+\frac{T_2}{\gamma_2}\right)-\lambda_n\bigg\}f_n(y)=0.
\end{multline}
To eliminate the term related to $y\partial_y$, we found that one needs to set
\begin{eqnarray}
T=\frac{k_B\left(\frac{T_1}{\gamma_1}+\frac{T_2}{\gamma_2}\right)}{k\left(\frac{1}{\gamma_1}+\frac{1}{\gamma_2}-\frac{2k_BT_1}{\gamma_1}\bm{i}\chi\right)}.
\end{eqnarray}
Thus, the above equation reduces to:
\begin{equation}
\left\{k_B\left(\frac{T_1}{\gamma_1}+\frac{T_2}{\gamma_2}\right)\frac{\partial^2}{\partial
y^2}+\bigg[-\frac{k^2\left(\frac{1}{\gamma_1}+\frac{1}{\gamma_2}-\frac{2k_BT_1}{\gamma_1}\bm{i}\chi\right)^2}{4k_B\left(\frac{T_1}{\gamma_1}+\frac{T_2}{\gamma_2}\right)}+k^2\left(\frac{k_BT_1}{\gamma_1}\left(\bm{i}\chi\right)^2-\frac{1}{\gamma_1}\bm{i}\chi\right)\bigg]y^2+\frac{k}{2}\left(\frac{1}{\gamma_1}+\frac{1}{\gamma_2}\right)-\lambda_n\right\}f_n(y)=0.
\end{equation}
Let us set $r_{1,2}=k/\gamma_{1,2}$,
$D_{1,2}=k_BT_{1,2}/\gamma_{1,2}$, then we have:
\begin{eqnarray}
\left\{\left(D_1+D_2\right)\frac{\partial^2}{\partial
y^2}+\left[-\frac{\left(r_1+r_2-2\bm{i}\chi
kD_1\right)^2}{4\left(D_1+D_2\right)}+\left(\bm{i}\chi\right)^2k^2D_1-\bm{i}\chi
kr_1\right]y^2+\frac{r_1+r_2}{2}-\lambda_n\right\}f_n(y)=0,
\end{eqnarray}
\begin{eqnarray}
\left\{\left(D_1+D_2\right)\frac{\partial^2}{\partial
y^2}-\frac{\left(r_1+r_2\right)^2}{4\left(D_1+D_2\right)}\left[1+\frac{4k\bm{i}\chi}{\left(r_1+r_2\right)^2}\left(D_2r_1-D_1r_2-\bm{i}\chi
kD_1D_2\right)\right]y^2+\frac{r_1+r_2}{2}-\lambda_n\right\}f_n(y)=0.
\end{eqnarray}
Then set $r=r_1+r_2$, $D=D_1+D_2$, $\theta=\sqrt{1+\frac{4k^2D_1D_2\chi}{r^2}(\bm{i}(1/T_2-1/T_2)-\chi)}$, we can simplify the above equation as
\begin{eqnarray}
\left\{D\frac{\partial^2}{\partial
y^2}-\frac{r^2}{4D}\theta^2y^2+\frac{r}{2}-\lambda_n\right\}f_n(y)=0
\end{eqnarray}
We further set $Y=\sqrt{\frac{r\theta}{2D}} y$ and $F_n(Y)=f_n(y)$, which finally leads to the eigen-problem of quantum harmonic oscillator's Schr\"{o}dinger's equation:
\begin{eqnarray}
\bigg\{\frac{\partial^2}{\partial
Y^2}-Y^2+\frac{1}{\theta}(1-\frac{2\lambda_n}{r})\bigg\}F_n(Y)=0,
\end{eqnarray}
This equation requires $\frac{1}{\theta}\left(1-\frac{2\lambda_n}{r}\right)=1+2n$, with $n=0,1,2,\ldots$. From any textbook of quantum mechanics which solves the Schr\"{o}dinger's equation of quantum harmonic oscillator, we know the eigenvalue:
\begin{eqnarray}
\lambda_n=\frac{r}{2}\left[1-\left(1+2n\right)\theta\right],
\end{eqnarray}
and the eigenfunction
\begin{eqnarray}
f_n(y)=e^{-\frac{r\theta}{4D}y^2}H_n(\sqrt{\frac{r\theta}{2D}}y),
\end{eqnarray}
where $H_n$ denotes the $n$-th order Hermite polynomials.

Since we already have $T=D/(r-2\bm{i}\chi kD_1)$, as a consequence, the ansatz of the solution now reads:
\begin{eqnarray}
z(y,\chi,t)=\sum_{n=0}^{\infty}C_ne^{\lambda_nt}e^{-\frac{r(1+\theta)-2\bm{i}\chi
kD_1}{4D}y^2}H_n(\sqrt{\frac{r\theta}{2D}}y)
\end{eqnarray}
Substitute it back to the twisted Fokker-Planck equation $\partial_t z=L_{\chi}z$, we have
\begin{eqnarray}
L_{\chi}
e^{-\frac{r(1+\theta)-2\bm{i}\chi
kD_1}{4D}y^2}H_n(\sqrt{\frac{r\theta}{2D}}y)=\lambda_ne^{-\frac{r(1+\theta)-2i\chi
kD_1}{4D}y^2}H_n(\sqrt{\frac{r\theta}{2D}}y),
\end{eqnarray}
such that $\lambda_n$ is indeed the eigenvalue of the operator $L_{\chi}$, the right eigenfunction for the operator $L_{\chi}$ is,
\begin{eqnarray}
\psi_n(y,\chi)=e^{-\frac{r(1+\theta)-2\bm{i}\chi
kD_1}{4D}y^2}H_n(\sqrt{\frac{r\theta}{2D}}y).
\end{eqnarray}
Straightforwardly, the corresponding left eigenfunction, which is bi-orthonormal with the right one, reads,
\begin{eqnarray}
\varphi_n(y,\chi)=\frac{1}{2^nn!}\sqrt{\frac{r\theta}{2\pi
D}}e^{\frac{r(1-\theta)-2\bm{i}\chi
kD_1}{4D}y^2}H_n(\sqrt{\frac{r\theta}{2D}}y),
\end{eqnarray}
This left eigenfunction of $L_{\chi}$, is just the right eigenfunction of the adjoint operator $L_{\chi}^+$, which has the property \cite{riskens}:
\begin{eqnarray}
L_{\chi}^+\varphi_n(y,\chi)=\lambda_n\varphi_n(y,\chi),
\end{eqnarray}
corresponding to
\begin{eqnarray}
L_{\chi}\psi_n(y,\chi)=\lambda_n\psi_n(y,\chi),
\end{eqnarray}
so that we have the scalar product
\begin{eqnarray}
\int^{+\infty}_{-\infty}dy\varphi_m(y,\chi)L_{\chi}\psi_n(y,\chi)=\int^{+\infty}_{-\infty}dy L_{\chi}^+\varphi_m(y,\chi)\psi_n(y,\chi)=\int^{+\infty}_{-\infty}dy \psi_n(y,\chi)L_{\chi}^+\varphi_m(y,\chi).
\end{eqnarray}
We also may thus normalize the functions according to
\begin{eqnarray}
\int^{+\infty}_{-\infty}dy\varphi_m(y,\chi)\psi_n(y,\chi)=\delta_{mn}
\end{eqnarray}
Under the adjoint operation ``+'' \footnote{We note here that, by the definition in this work, the adjoint operation ``+'' is similar to the transpose operation in matrix, but not the conjugate transpose.}, one can easily find that
\begin{eqnarray}
y\rightarrow y, \frac{\partial}{\partial y}\rightarrow-\frac{\partial}{\partial y},
y\frac{\partial}{\partial y}\rightarrow-\frac{\partial}{\partial y}y=-\textbf{1}-y\frac{\partial}{\partial y}.
\end{eqnarray}
Therefore, from Eq. (19), we have
\begin{equation}
L_{\chi}^+=k_B\left(\frac{T_1}{\gamma_1}+\frac{T_2}{\gamma_2}\right)\frac{\partial^2}{\partial
y^2}
-k\left(\frac{1}{\gamma_1}+\frac{1}{\gamma_2}-\frac{2k_BT_1}{\gamma_1}\bm{i}\chi\right)y\frac{\partial}{\partial
y}+k^2\left(\frac{k_BT_1}{\gamma_1}\left(\bm{i}\chi\right)^2-\frac{1}{\gamma_1}\bm{i}\chi\right)y^2+k\frac{k_BT_1}{\gamma_1}\bm{i}\chi.
\end{equation}
Then, following the similar procedure, one is able to arrive at the left eigenfunction Eq. (35).

Behaviors in the long time limit are of our central interest. They are
governed by the ground state of the twisted Fokker-Planck operator $L_{\chi}$, of which the eigenvalue, $\lambda_0(\chi)$, possesses the least negative real part. In other words, for time-independent $L_{\chi}$, $\lim_{t\rightarrow\infty}Z(\chi,t)\sim e^{\lambda_0(\chi)t}$ and in turn $\lim_{t\rightarrow\infty}\langle\langle Q^n\rangle\rangle/t=\partial^n_{\bm{i}\chi} \lambda_0(\chi)|_{\chi=0}$. If the system is under time-dependent modulation such that the Fokker-Planck operator is time-dependent $L_{\chi}(t)$, then as discussed in the main text, the full counting statistics rely on the instantaneous right and left ground-state $\psi_0(y,\chi,t)$ and $\varphi_0(y,\chi,t)$.
Therefore, we only needs the information about the ground state. From Eqs. (30, 34, 35), we finally have the instantaneous ground state:
\begin{align*}
&\lambda_0(\chi,t)=\frac{r}{2}\left(1-\theta\right)=\frac{1}{2}\left(\frac{k}{\gamma_1}+\frac{k}{\gamma_2}-
\sqrt{(\frac{k}{\gamma_1}+\frac{k}{\gamma_2})^2+\bm{i}\chi
\frac{4k^2}{\gamma_1\gamma_2}
\big(k_BT_2-k_BT_1-\bm{i}\chi k_B^2T_1T_2\big)}\right), \\
&\psi_0(y,\chi,t)=\exp\left({-\frac{r+r\theta-2\bm{i}\chi kD_1}{4D}y^2}\right), \quad
\varphi_0(y,\chi,t)=\sqrt{\frac{r\theta}{2\pi D}}\exp\left({\frac{r-r\theta-2\bm{i}\chi
kD_1}{4D}y^2}\right).
\end{align*}
Similar solutions in a Fokker-Planck equation without the counting parameter $\chi$ were discussed in \cite{riskens,peters}.
Similar results about the eigenvalue and the right eigenfunction, but with the left eigenfunction absent, were given by \cite{Visco2s}.

\subsection{Condition for ``adiabatic''}
From Eq. (3), we write down the first three term for the sake of clarity:
\begin{eqnarray}
z(y,\chi,t)&=&\widetilde{C}_0e^{\lambda_0t}+\widetilde{C}_1e^{\lambda_1t}+\widetilde{C}_2e^{\lambda_2t}+... \nonumber \\
&=&\widetilde{C}_0e^{\lambda_0t}\left(1+\frac{\widetilde{C}_1}{\widetilde{C}_0}e^{-(\lambda_0-\lambda_1)t}+\frac{\widetilde{C}_2}{\widetilde{C}_0}e^{-(\lambda_0-\lambda_2)t}+...\right),
\end{eqnarray}
where $\widetilde{C}_n$ are some unimportant coefficients here. Clearly, $1/(\lambda_0(\chi)-\lambda_n(\chi))$ depicts the characteristic relaxation time of the $n$-th mode to the ground state. Let us set the counting parameter $\chi=0$ to look into spectrum of the operator $L_{\chi=0}$ in the real physical space. From Eq. (30), we can see that the spectrum is ordered by $n$ and the eigenvalue of the ground state $\lambda_0(\chi=0)=0$. This is apparent, since it corresponds to the long time steady state of the evolution. So, the rate of relaxation to the steady state is determined by the inverse energy gap $1/(\lambda_0(\chi=0)-\lambda_1(\chi=0))$. As a consequence, the system's characteristic time scale of relaxation is given as
\begin{equation}
T_c=\left.\frac{1}{\lambda_0(\chi)-\lambda_1(\chi)}\right|_{\chi=0}=\frac{\gamma_1\gamma_2}{k(\gamma_1+\gamma_2)},
\end{equation}
where $\lambda_1(\chi=0)=k/\gamma_1+k/\gamma_2$ has the physics meaning of the effective damping rate. Therefore, given the system, which is already in its steady state after a long time evolution, as long as we drive it slowly such that the driving period $T_p\gg T_c$, the system can always dwell in its steady state (ground state). In other words, $T_p\gg T_c$ is right the so-called ``adiabatic'' condition.

\subsection{Geometric phase contribution in cumulant generating functions}
We assume the system is already at its steady state, or say ground state. And then, time-dependent modulations are imposed on the system adiabatically. Thus, we can make the ansatz
$z(y,\chi,t)=C_0(t)e^{\int_0^t\lambda_0(\chi,t')dt'}\psi_0(y,\chi,t)$, and substitute it
into the time-dependent twisted Fokker-Planck equation $\partial_t z(y,\chi,t)=L_{\chi}(t)z(y,\chi,t)$. We then obtain
\begin{equation}
\dot{C}_0(t)e^{\int_0^t\lambda_0(\chi,t')dt'}\psi_0(y,\chi,t)+C_0(t)
e^{\int_0^t\lambda_0(\chi,t')dt'}\dot{\psi}_0(y,\chi,t)=0.
\end{equation}
Utilizing the bi-orthonormal condition
$\int^{\infty}_{-\infty} dy\varphi_m(y,\chi,t)\psi_n(y,\chi,t)=\delta_{mn}$,
we left multiply $\varphi_0$, and
do integral over $y$, such that:
\begin{equation}
\dot{C}_0(t)=-C_0(t)\int^{\infty}_{-\infty}
dy\varphi_0(y,\chi,t)\dot{\psi}_0(y,\chi,t).
\end{equation}
Therefore, in the adiabatic limit, we obtain
\begin{equation}
C_0(t)=C_0(0)e^{-\int_0^tdt'\int^{\infty}_{-\infty}
dy\varphi_0(y,\chi,t')\dot{\psi}_0(y,\chi,t')}.
\end{equation}
Consequently, the characteristic function finally reads
\begin{equation}
Z(\chi,t)= \left[\int^{\infty}_{-\infty}  dy C_0(0)\psi_0(y,\chi,t)\right]\exp\left[{\int_0^t\lambda_0(\chi,t')dt'}\right]
\exp\left[{-\int_0^tdt'\int^{\infty}_{-\infty}  dy\varphi_0(y,\chi,t')\dot{\psi}_0(y,\chi,t')}\right].
\end{equation}
The first exponent, the time integral of the instantaneous ground-state eigenvalue, is an analog of the dynamic phase. While the second additional exponent resulting from the time-evolving of ground eignstates is an analog of the geometric phase. Different from the conventional phase of the wave function in quantum mechanics, here the ``phase'' refers to the cumulant generating function in the exponent of the characteristic function, which will contribute to the full counting statistics of the quantities of interest. Similar geometric phase contribution to the generating function so as to the full counting statistics is firstly discovered by Sinitsyn and Nemenman in discrete chemical kinetics \cite{Sinitsyns}.

Successively, the cumulant generating function of adiabatically driven systems can be separated into two parts -- the dynamic phase contribution and the geometric phase contribution:
$G(\chi)=\lim_{t\rightarrow\infty}\frac{1}{t}\ln Z(\chi,t)=G_{\mathrm{dyn}}+G_{\mathrm{geom}}$, with
\begin{align}
G_{\mathrm{dyn}}&=\frac{1}{T_p}\int^{T_p}_0dt\lambda_0(\chi,t), \\
G_{\mathrm{geom}}&=-\frac{1}{T_p}\int^{T_p}_0dt\int^{\infty}_{-\infty}
dy\varphi_0(y,\chi,t)\dot{\psi}_0(y,\chi,t).
\end{align}
Note that here the contribution of initial conditions $t^{-1}\ln\big[\int^{\infty}_{-\infty} dy
C_0(0)\psi_0(y,\chi,t)\big]$ is assumed negligible in the long time limit. The dynamic phase contribution survives whenever system parameters are static, or experience single or multiple modulations. While the existence of the geometric phase contribution requires at least two parameter modulations. For the case of periodically driving $(u_1(t), u_2(t))$, which could be chosen from $k, \gamma_j, T_j$, using Stokes theorem, we have
\begin{eqnarray}
G_{\mathrm{geom}}=-\frac{1}{T_p}\iint_{u_1u_2}du_1du_2
\mathcal{F}_{u_1u_2}(\chi),
\end{eqnarray}
where the subscript $u_1u_2$ denotes the integral area enclosed by the modulating contour of $(u_1(t), u_2(t))$, and
\begin{eqnarray}
\mathcal{F}_{u_1u_2}(\chi)\equiv\int^{\infty}_{-\infty} dy \left[\frac{\partial\varphi_0}{\partial u_1}\frac{\partial\psi_0}{\partial u_2}-\frac{\partial\varphi_0}{\partial u_2}\frac{\partial\psi_0}{\partial u_1}\right]
\end{eqnarray}
is a classical analog of the quantum mechanical Berry curvature. Different from the curvature usually defined for discrete Hilbert space of matrix-like Hamiltonian, the curvature here is in the continuous function space, indicated by the integral over infinity. Clearly, the curvature $\mathcal{F}_{u_1u_2}(\chi)$ is also of pure geometric origin, since it is independent of the modulation speed.

\end{widetext}


\begin{thebibliography}{01}

\bibitem{TDHT1} D. Segal and A. Nitzan, Phys. Rev. E {\bf73}, 026109 (2006); D. Segal, Phys. Rev. Lett. {\bf101}, 260601 (2008).
\bibitem{TDHT3} R. Marathe, A. M. Jayannavar, and A. Dhar, Phys. Rev. E {\bf75}, 030103(R) (2007)

\bibitem{TDHT2} N. Li, P. H\"anggi, and B. Li, Europhys. Lett. {\bf 84}, 40009 (2008)
\bibitem{TDHT22} N. Li, F. Zhan, P. H\"anggi, and B. Li, Phys. Rev. E {\bf80}, 011125 (2009)
\bibitem{TDHT23} J. Ren and B. Li, Phys. Rev. E {\bf81}, 021111 (2010).

\bibitem{TDHT32} A. Dhar, O. Narayan, A. Kundu, and K. Saito, Phys. Rev. E {\bf83}, 011101 (2011).

\bibitem{TDHT4} S. Zhang, J. Ren, and B. Li, Phys. Rev. E {\bf84}, 031122 (2011);
B. Q. Ai, D. He, and B. Hu, Phys. Rev. E {\bf 81}, 031124 (2010).

\bibitem{TDHT5} E. C. Cuansing and J.-S. Wang Phys. Rev. E 82, 021116 (2010); B. K. Agarwalla, J.-S. Wang, and B. Li, Phys. Rev. E {\bf 84}, 041115 (2011).

\bibitem{Ren} J. Ren, P. H\"anggi, and B. Li, Phys. Rev. Lett. {\bf 104}, 170601 (2010).

\bibitem{Li_review} N. Li, \emph{et al.} arXiv:1108.6120 (accepted by Rev. Mod. Phys.)

\bibitem{GePh} D. Xiao, M.-C. Chang, and Q. Niu, Rev. Mod. Phys. {\bf82}, 1959 (2010).
\bibitem{GePh1} E. Prodan and C. Prodan, Phys. Rev. Lett. {\bf103}, 248101 (2009).
\bibitem{GePh2} J.-T. L\"u, M. Brandbyge, and P. Hedeg{\aa}rd, Nano Lett. {\bf10}, 1657 (2010).
\bibitem{GePh3} L. Zhang, J. Ren, J.-S. Wang, and B. Li, Phys. Rev. Lett. {\bf105}, 225901 (2010).

\bibitem{Parrondo1998} J. M. R. Parrondo, Phys. Rev. E {\bf57}, 7297 (1998).
\bibitem{Shi2002} Y. Shi and Q. Niu, Europhys. Lett. {\bf59}, 324 (2002).
\bibitem{Astumian} R. D. Astumian and P. H\"anggi, Phys. Today {\bf55}, 33 (2002); R. D. Astumian, Proc. Natl. Acad. Sci. U.S.A. {\bf104}, 19715 (2007); S. Rahav, J. Horowitz, and C. Jarzynski, Phys. Rev. Lett. {\bf101}, 140602 (2008).

\bibitem{Sinitsyn} N. A. Sinitsyn and I. Nemenman, Europhys. Lett. {\bf 77}, 58001 (2007); Phys. Rev. Lett. {\bf99}, 220408 (2007).
\bibitem{Sinitsyn_review} N. A. Sinitsyn, J. Phys. A: Math. Theor. {\bf42}, 193001 (2009).

\bibitem{ADhar_review} A. Dhar, Adv. Phys. 57, 457 (2008); S. Lepri, R. Livi, A. Politi, Phys. Rep. {\bf377}, 1 (2003).

\bibitem{ratchet} J. M. R. Parrondo and P. Espa\~nol, Am. J. Phys. {\bf 64}, 1125 (1996).

\bibitem{Purcell} E. M. Purcell, Am. J. Phys. \textbf{45}, 3 (1977).

\bibitem{FP} N. G. van Kampen, {\it Stochastic Processes in Physics and Chemistry} (Elsevier, North-Holland, 2007).

\bibitem{Visco1} P. Visco, A. Puglisi, A. Barrat, E. Trizac, F. van Wijland, J. Stat. Phys. {\bf 125}, 533 (2006).

\bibitem{LD_harmonic} A. Kundu, S. Sabhapandit, and A. Dhar, J. Stat. Mech. {\bf P03007}, (2011).

\bibitem{supply} See supplymentary material.


\bibitem{Berry} M. V. Berry, Proc. R. Soc. Lond. A {\bf 392}, 45 (1984).
\bibitem{Garrison88} J. G. Garrison and E. M. Wright, Phys. Lett. A {\bf128}, 177
(1988).

\bibitem{GUC} V. Y. Chernyak, M. Chertkov, and N. A. Sinitsyn, J. Stat. Mech. {\bf P09006}, (2011).
\bibitem{Horowitz} J. M. Horowitz and C. Jarzynski, J. Stat. Phys. {\bf136}, 917 (2009); Jie Ren, V. Y. Chernyak, and N. A. Sinitsyn, J. Stat. Mech. {\bf P05011}, (2011).

\bibitem{Temp_viscosity} J. C. Maxwell, {\it Maxwell on Molecules and Gases}, Editor E. Garber, S. G. Brush and C. W. F. Everitt, (The MIT Press, 1986).

\bibitem{Li2007} N. Li and B. Li, Europhys. Lett. {\bf 78}, 34001 (2007).

\bibitem{SCPT} D. He, S. Buyukdagli, and B. Hu, Phys. Rev. E {\bf 78}, 061103 (2008).

\bibitem{wangEPJB} J.-S. Wang, J. Wang, and J. T. L\"u, Eur. Phys. J. B. {\bf 62}, 381 (2008).

\bibitem{RC} R. van Zon, S. Ciliberto, and E. G. D. Cohen, Phys. Rev. Lett. {\bf 92}, 130601 (2004).

\bibitem{FTexp} N. Garnier and S. Ciliberto, Phys. Rev. E {\bf 71}, 060101(R) (2005).

\bibitem{Diodeexp} F. Tao, W. Chen, W. Xu, J. Pan, and S. Du, Phys. Rev. E {\bf 83}, 056605 (2011).

\end{thebibliography}

\begin{thebibliography}{01}
\bibitem{riskens} H. Risken, {\it The Fokker-Planck Equation: Methods of Solution and Applications}, (Springer-Verlag, New York) (1996).
\bibitem{peters} P. H\"anggi, H. Grabert, P. Talkner, and H. Thomas, Phys. Lett. A {\bf 29}, 371 (1984).
\bibitem{Visco2s} P. Visco, J. Stat. Mech. {\bf P06006}, (2006).
\bibitem{Sinitsyns} N. A. Sinitsyn and I. Nemenman, Europhys. Lett. {\bf 77}, 58001 (2007); Phys. Rev. Lett. {\bf99}, 220408 (2007).
\end{thebibliography}
\end{document}